\documentclass[aps,twocolumn,prd,nofootinbib]{revtex4}%

\usepackage{graphicx}
\usepackage{amsmath,amssymb}
\usepackage{hyperref}
\usepackage[usenames]{color}

\hypersetup{
    colorlinks=true,
    linkcolor=red,
    citecolor=blue,
}

\def\be{\begin{equation}}
\def\ee{\end{equation}}
\def\ba{\begin{eqnarray}}
\def\ea{\end{eqnarray}}

\begin{document}

\title{Probing Dark Energy Dynamics from Current and Future Cosmological Observations }

\author{Gong-Bo Zhao$^{1,2}$}
\email{Gong-bo.Zhao@port.ac.uk}
\author{Xinmin Zhang$^{3,4}$}
\email{xmzhang@ihep.ac.cn}

\affiliation{ $^1$ Institute of Cosmology and Gravitation, Dennis
Sciama Building, Burnaby Road, Portsmouth, PO1 3FX, United Kingdom
\\
$^2$Department of Physics, Simon Fraser University, Burnaby, BC, V5A 1S6, Canada \\
$^3$Theoretical Physics Division, Institute of High Energy Physics,
Chinese Academy of Sciences, P.O.Box 918-4, Beijing 100049,
P.R.China\\
$^4$Theoretical Physics Center for Science Facilities (TPCSF),
Chinese Academy of Sciences, P.R.China }

\begin{abstract}

We report the constraints on the dark energy equation-of-state
$w(z)$ using the latest `Constitution' SNe sample combined with the
WMAP5 and SDSS data. Assuming a flat universe, and utilizing the
localized principal component analysis and the model selection
criteria, we find that the $\Lambda$CDM model is generally
consistent with the current data, yet there exists weak hint of the
possible dynamics of dark energy. In particular, a model predicting
$w(z)<-1$ at $z\in[0.25,0.5)$ and $w(z)>-1$ at $z\in[0.5,0.75)$,
which means that $w(z)$ crosses $-1$ in the range of
$z\in[0.25,0.75)$, is mildly favored at 95\% confidence level. Given
the best fit model for current data as a fiducial model, we make
future forecast from the joint data sets of JDEM, Planck and LSST,
and we find that the future surveys can reduce the error bars on the
$w$ bins by roughly a factor of $10$ for a 5-$w$-bin model.

\end{abstract}

\maketitle

\section{Introduction}

The apparent acceleration of the universe discovered at the end of
last century remains an enigma~\cite{Riess98,Perl99}, and much
effort has been made to find consistent explanation. This includes
the modification or even the `reinvention' of the Einstein gravity
on large
scales~\cite{Starobinsky:1980te,Capozziello:2003tk,Carroll:2003wy,Starobinsky:2007hu,Nojiri:2008nk,Boisseau:2000pr,Khoury:2003aq,Dvali:2000hr,Dvali:2007kt},
and also the addition of the exotic energy budget -- \emph{Dark
Energy} (DE) term to the right hand side of the Einstein equation.
The \emph{Equation-of-State} (EoS) $w$ of dark energy, defined as
the ratio of pressure and energy density, is usually used to
classify different DE models. For example, the $w$ of the simplest
DE candidate, the vacuum energy, is a constant of $-1$, while $w$ is
generally considered as a function of redshift $z$ for models
predicting dark energy dynamics such as
quintessence~\cite{Ratra:1987rm}, phantom~\cite{Caldwell:1999ew},
quintom~\cite{Feng:2004ad} and so forth.

It is true that we are still ignorant about the nature of dark
energy, yet the accumulating high precision observational data of
\emph{Supernova Type Ia} (SNe), \emph{Cosmic Microwave Background}
(CMB) and \emph{Large Scale Structure} (LSS) make it possible to
study dark energy phenomenologically, i.e., confronting the numerous
dark energy models to data and narrowing down the dark energy
candidates by falsifying part of the models, which might be the best
we can do to approach the truth of dark energy. In this data-driven
investigation, in most cases one needs to parameterize $w(z)$ in the
first place, i.e. assume an \emph{ad hoc} functional form of $w(z)$,
then fit the related parameters to data, reconstruct $w(z)$ and make
statements according to the result of the
reconstruction~\cite{DE_func_recons}. Since the result may more or
less depend on the assumed form of $w(z)$, and to minimize the
artifacts, one needs to choose the parametrization with care -- it
should be physically motivated, and statistically sound, i.e., using
the least number of free parameters to obtain the maximum
generality.

The first example of the viable parametrization is the widely used
CPL parametrization~\cite{CPL:CP,CPL:L}, \be\label{eq:CPL}
w(a)=w_{0}+w_{a}\frac{z}{1+z}\ee where $w_0$ and $w_a$ are free
parameters. It is widely used since it has a simple form, clear
interpretation -- $w_0$ is the EoS today and $w_a$ denotes the
derivative with respective to the scale factor $a$, thus a dark
energy dynamics indicator -- and a small number of free parameters.
However, the simplicity of the form inhabits the CPL parametrization
to describe the models whose EoS deviates significantly from the
linear function of $a$, for example, a $w(z)$ with
oscillations~\cite{osc}, or with other local features~\cite{bump}.

One alternative is to approximate $w(z)$ using the piecewise
constant bins~\cite{Huterer:2002hy}, which is much more general than
the functional parametrizations. This generality allows for the
high-resolution temporal reconstruction of $w(z)$, based on which
one can make further model-independent studies using a
\emph{Principal Component Analysis} (PCA) method
\cite{Huterer:2002hy,Hu:2002rm,Crittenden:2005wj,dePutter:2007kf,Zhao:2007ew,Serra:2009yp,dePutter:2008bh,Zhao:2009fn}.
For example, it is possible to know how many EoS parameters can be
well constrained by current/future data regardless of the forms of
parametrization; where is (are) the sweet spot(s) (the redshift
where the error on $w(z)$ gets minimized); and we can even construct
the uncorrelated bands to probe dark energy dynamics more explicitly
from the cosmological
observations~\cite{LPCA1,LPCA2,dePutter:2007kf,Zhao:2007ew}
\footnote{Aside from PCA, other less model-dependent methods have
been developed, such as smoothing the SN data to derive
$w(z)$~\cite{Shafieloo:2005nd}, constraining the binned dark energy
density~\cite{Wang:2007mza}, etc.}. Given the constraints on $w(z)$
from the current data, it is useful to know to what extent the
future surveys will tighten the constraints. In this paper, we will
focus on the \emph{Uncorrelated Band-power Estimates} (UBE) of
$w(z)$ from the current and the simulated future data.

In the next section, we will describe the method and data we use in
detail and in section III we will present our main results, and then
we will finish with the summary and discussion.

\section{Methodology and Data}\label{method}
\subsection{Constraining dark energy from current observations}
To fit to data, we parametrize our universe as: \be
\label{eq:paratriz} {\bf P} \equiv (\omega_{b}, \omega_{c},
\Theta_{s}, \tau, n_s, A_s, \mathcal{X}) \ee where
$\omega_{b}\equiv\Omega_{b}h^{2}$ and
$\omega_{c}\equiv\Omega_{c}h^{2}$ are the physical baryon and cold
dark matter densities relative to the critical density respectively,
$\Theta_{s}$ stands for the ratio (multiplied by 100) of the sound
horizon to the angular diameter distance at decoupling, $\tau$
denotes the optical depth to re-ionization, and $n_s, A_s$ are the
primordial power spectrum index and amplitude, respectively. We
assume a flat universe throughout.

The dark energy EoS parameters are denoted by $\mathcal{X}$, and we
consider the following two kinds of parametrizations in this work,
\be \label{paraw} w(z)= \left\{
    \begin{array}{ll}
    {\rm Sum~of~the~tanh~bins}, & \hbox{$\mathcal{X}_{\rm I}~~~=
    \{w_i\}$;} \\
    \hbox{$w_0+w_a\cdot{z}/(1+z)$}, & \hbox{$\mathcal{X}_{\rm II}~~= \{w_0, w_a\}$;} \\
    \end{array}
\right. \ee where $\mathcal{X}_{\rm I}$ and $\mathcal{X}_{\rm II}$
are the collections of the bin parameters for dark energy equation
of state (see explanations below) and the CPL parameters
respectively, and the constraints on these parameters allow us to
reconstruct the evolution history of $w(z)$, which might encode the
dark energy dynamics.

For dark energy parametrization $\mathcal{X}_{\rm I}$, we
approximate $w(z)$ using the sum of $N$ piecewise constant bins
$w_i$ localizing in redshift and vary them to fit data. The simplest
realization of this binning is to use step
functions~\cite{Huterer:2002hy, LPCA1, LPCA2,dePutter:2007kf}.
However, the resulting discontinuity in $w(z)$ makes it difficult to
handle dark energy perturbations, which depends on the time
derivative of $w(z)$ and plays a crucial role in the parameter
estimation, and should be treated in a consistent
way~\cite{Weller:2003hw,Zhao:2005vj,Fang:2008sn}. To solve the
problem of discontinuity, one can use the smooth and differentiable
functions such as the cubic spline functions
\cite{Zhao:2007ew,Serra:2009yp}, and the hyperbolic tangent
functions \cite{Crittenden:2005wj} for the binning. Here we follow
the latter and parameterize $w(z)$ as, \be\label{Eq:w_z}
w(z)=\sum_{i=1}^{N-1} \frac{(w_{i+1}-w_{i})}{2}\Big[1+{\rm
tanh}\Big(\frac{z-z_{i+1}}{\xi}\Big)\Big] +w_{1} \ee where $w_i$
denotes the value of EoS in the $i$th bin, and $z_i, z_{i+1}$ stand
for the endpoints of the $i$th bin. Also note that $\xi$ is the
transition width of two neighboring bins, and is set to $5\%$ of the
bin width. We have numerically checked that the final result is
largely independent of $\xi$ as long as it describes a sharp, but
numerically stable transition. For a total of $N$-bin
parametrization, we arrange the first $N-1$ bins to be evenly spaced
at low redshifts ($z\leq1$), and use one wide bin to model $w(z)$ at
$z>1$ since dark energy becomes less and less important as redshift
increases, and it has been found that there is no hope to resolve
dark energy dynamics, if any, beyond redshift one even using the
future \emph{Joint Dark Energy Mission} (JDEM) survey
\cite{Huterer:2002hy,JDEM}. In our numerical analysis, we let all
the $N$ bins (including the high-$z$ bin) float in all cases, thus
our constraint on $w(z)$ at low-$z$ is more conservative than that
with fixed high-$z$ bin. Note that in our notation, $N=1$
corresponds to the $w$CDM model where $w$ is a constant regardless
of redshift, and $N=0$ stands for $\Lambda$CDM model.

It is true that the larger $N$ we use, the higher temporal
resolution we can obtain. However, the result will be severely
diluted for large $N$. This is not only due to the weakness of
current data, but also to the huge degeneracy introduced. Therefore
we need a criterion to determine the optimal number of bins, so that
we can detect the main features from data by using the minimal
number of bins. This is an issue of Occam's razor, i.e., we don't
want to introduce unnecessary parameters. To optimize the
\emph{Goodness of Fit} (GoF), we search for an optimal $N$ in the
range $N\in[0,10]$ based on the model selection criteria which we
will describe in detail in Sec. \ref{result}. Then $w(z)$ can be
reconstructed from data.

 However, the
interpretation of the reconstructed $w(z)$ might be obscured by the
correlations among the dark energy bins. To eliminate this
ambiguity, one seeks for a linear transformation $\mathcal{W}$ to
rotate the original parameter vector $\mathcal{X}$ into a new
parameter vector $\mathcal{Q}=\mathcal{W}\mathcal{X}$ so that the
resultant new parameters, denoted by the $q$'s, are physically and
statistically meaningful, namely, they directly relate to the $w$
bins and have uncorrelated errors.

This is an eigenvalue/vector problem, and can be solved by applying
the linear manipulations on the covariance matrix $\mathbf{C}$ of
the $w$ bins (after marginalizing over the other cosmological
parameters): \be\label{eq:C_X1} \mathbf{C_{\mathcal{X_{\rm I}}}} =
(w_i-\langle{w_i}\rangle)(w_j-\langle{w_j}\rangle)^T
             = \langle\mathcal{X_{\rm I}}\mathcal{X_{\rm I}}^T\rangle~\ee
We then diagonalize the Fisher matrix $\mathbf{F} \equiv
\mathbf{C}^{-1}$ so that
 $\mathbf{F} = \mathbf{O}^T \mathbf{\Lambda}\mathbf{O}
 = \mathcal{W}^T\mathcal{W}$. Here $\mathbf{O}$ is the orthogonal
 matrix, $\mathbf{\Lambda}$ is diagonal and $\mathcal{W}=\mathbf{F}^{1/2}$. Then the resulting transformation
matrix $\mathcal{W}$, obtained by absorbing $\mathbf{\Lambda}^{1/2}$
into $\mathbf{O}$, can make the new parameters $q$'s uncorrelated
and the weights (rows of $\mathcal{W}$) are nearly positive definite
and localized in redshift, which means that $q_i$ has a one-to-one
correspondence to $w_i$ (and close to $w_i$), making the
interpretation more transparent
\cite{LPCA1,LPCA2,dePutter:2007kf,Zhao:2007ew}. To make $q(z)=-1$
stand for $\Lambda$CDM, we re-scale $\mathcal{W}$ so that its rows
sum up to unity, i.e.
$\mathcal{W}_{ij}=F_{ij}^{1/2}/(\sum_k{F_{ik}^{1/2}})$. Note,
however, that the normalization is arbitrary and the physical
inferences do not depend on this normalization as both the
parameters $q$'s and their values for a particular theoretical model
change consistently \cite{Zhao:2007ew}. Before normalization, the
variances of the $q$'s are unity since \be\label{eq:C_Q}
\mathbf{C}_\mathcal{Q}=\langle{\mathcal{Q}\mathcal{Q}^T}\rangle=\mathcal{W}\langle\mathcal{X_{\rm
I}}\mathcal{X_{\rm
I}}^T\rangle\mathcal{W}^T=\mathbf{F}^{1/2}\mathbf{C}{\mathbf{F}^{1/2}}=\mathbf{I}\ee
While after normalization, the covariance matrix becomes
$\langle\Delta{q_i}\Delta{q_j}\rangle=\delta_{ij}/(\sum_a{F_{ia}^{1/2}}\sum_b{F_{jb}^{1/2}})$.
This prescription is called the Uncorrelated Band-power Estimate, or
\emph{Localized Principal Component Analysis} (LPCA).

Parametrization $\mathcal{X}_{\rm II}$ is a widely used functional
form for dark energy EoS, but it needs to be tested whether this
linear function of scale factor is general enough to describe dark
energy for the current data. This can be done by reconstructing
$w(z)$ from parametrizations $\mathcal{X}_{\rm I}$ and
$\mathcal{X}_{\rm II}$ respectively, and then making a direct
comparison.

Given the set of cosmological parameters ${\bf P}$ in
Eq~(\ref{eq:paratriz}), we calculate the observables including the
luminosity distance, CMB and matter power spectra, and the cosmic
age using our modified version of {\tt
CAMB}\footnote{http://camb.info/}, which is able to calculate all
the necessary observables for an arbitrary $w(z)$. We pay particular
attention to the Dark Energy Perturbations (DEP), especially when
$w(z)$ crosses $-1$. Our calculation for DEP for general dark energy
models is based on the work presented in Ref.~\cite{Zhao:2005vj}. We
then fit to SNe, CMB and LSS observations using a modified version
of the Markov Chain Monte Carlo (MCMC) package {\tt
CosmoMC}\footnote{http://cosmologist.info/cosmomc/}\cite{CosmoMC}
based on the Bayesian statistics.
\begin{figure*}[tbp]
\includegraphics[scale=3.5]{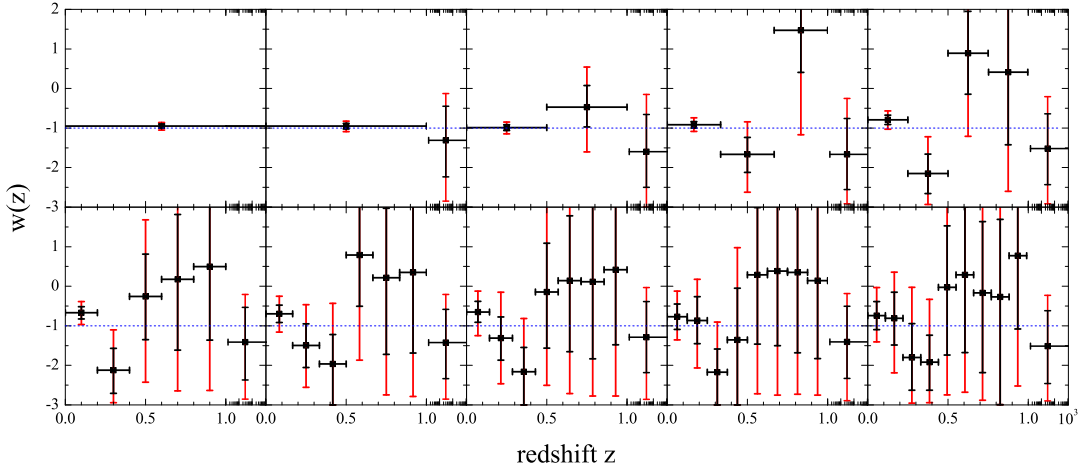}
\caption{ The correlated binned estimates of the EoS from current
observational data. The black and red error bars show $1\sigma$ and
$2\sigma$ uncertainties respectively. The blue dashed line shows
$\Lambda$CDM model. }\label{fig:wz}
\end{figure*}

The supernova data we use are the recently released ``Constitution"
SALT sample \cite{Hicken:2009dk}, and we marginalize over the
nuisance parameter, which is basically the calibration uncertainty
in measuring the supernova intrinsic magnitude, in the likelihood
calculation. For CMB, we use the WMAP five-year data including the
temperature and polarization power spectra~\cite{WMAP5}, and
calculate the likelihood using the routine supplied by the WMAP
team\footnote{http://lambda.gsfc.nasa.gov/}. For the LSS
information, we use the Sloan Digital Sky Survey (SDSS) Luminous Red
Galaxy (LRG) sample \cite{Tegmark:2006az}, and marginalize over the
bias parameter. Furthermore, we impose the $1-\sigma$ Gaussian
priors on the Hubble parameter and baryon density of $h=0.72\pm0.08$
and $\Omega_{b}h^{2}=0.022\pm0.002$ from the measurements of Hubble
Space Telescope (HST) \cite{HST} and Big Bang Nucleosynthesis
\cite{BBN} respectively, and a tophat prior on the cosmic age of 10
Gyr $< t_0 <$ 20 Gyr. The total likelihood is taken to be the
products of the separate likelihoods $\mathcal{L}$ of each dataset
we used, thus the total $\chi^2$ is the sum of separate $\chi^2$
from individual observations plus that from the priors if we define
$\chi^2 \equiv -2 \log {\mathcal{L}}$.

\begin{figure}
\includegraphics[scale=2]{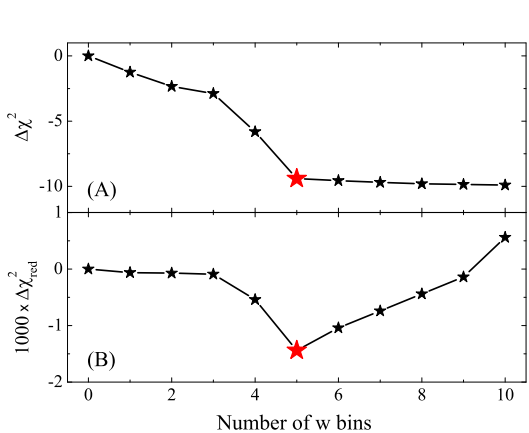}
\caption{The improved $\chi^2$ and reduced $\chi^2$ w.r.t. the
$\Lambda$CDM model as a function of the number of $w$ bins. The big
red stars illustrate the optimal model.}\label{fig:chi2}
\end{figure}

\begin{figure}
\includegraphics[scale=2]{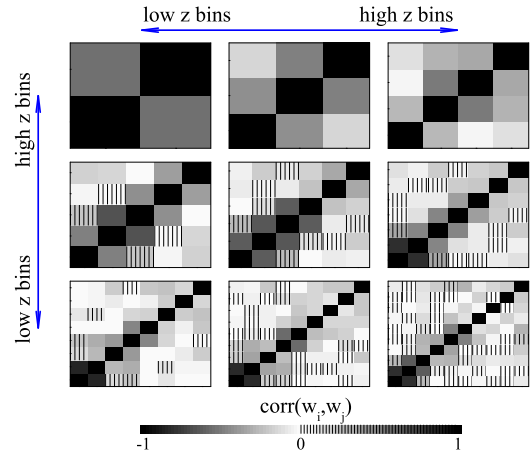}
\caption{The correlation matrices for the $w$ bins for different
binnings. From upper left to lower right, we show the correlation
matrices for $N=2$ to $N=10$ binnings (high-$z$ bin inclusive in all
cases, as stated in the main text).}\label{fig:corr}
\end{figure}

\subsection{Future Forecast}

Given the current constraints on dark energy, it is useful to know
quantitively how future surveys can improve the constraints.
Therefore we choose the best fit $w(z)$ from current data as a
fiducial model, and make a forecast from the surveys of JDEM,
Planck~\cite{planck} and \emph{Large Synoptic Survey Telescope}
(LSST)~\cite{LSST} by employing a standard Fisher Matrix technique
following~\cite{Zhao:2008bn}.

\subsubsection{Observables and Fisher Matrices} \label{observables}

Besides the luminosity distance and the CMB power spectra simulated
for JDEM and Planck respectively, we include the tomographic
observables including the spectra of the \emph{Galaxy Counts}(GC)
and \emph{Weak Lensing}(WL) from LSST, and all the possible
cross-correlations with CMB. Mathematically, all the tomographic
observables we use can be summarized as
\begin{equation} C_\ell^{XY}= 4\pi \int \frac{dk}{k} \Delta_{\cal
R}^{2} I_{\ell}^X(k) I_{\ell}^Y(k), \label{eq:gen}
\end{equation} where $\Delta_{\cal R}^{2}$ is the
primordial curvature power spectrum and $I_{\ell}^{X,Y}(k)$ denotes
the angular transfer functions. Here $X,Y\in[T,E,G_i,\epsilon_j]$,
where $T,E,G_i$ and $\epsilon_j$ illustrate the CMB temperature,
$E$-mode polarization, the $i$th redshift bin for galaxy counts and
the $j$th redshift bin for weak lensing shear respectively. In other
words, we consider all the possible cross-correlations among CMB, GC
and WL.

Given the specifications of the proposed future surveys, the tool of
Fisher matrix~\cite{Fisher} enables us to quickly estimate the
errors on the cosmological parameters around the fiducial values.
For zero-mean Gaussian-distributed observables, such as
$C^{XY}_\ell$, the Fisher matrix is given by \be F_{\alpha\beta} =
f_{\rm sky} \sum_{\ell=\ell_{\rm min}}^{\ell_{\rm max}}\frac{2\ell +
1}{2} {\rm Tr}\left( \frac{\partial {\bf C_\ell}}{\partial p_\alpha}
{\bf \tilde{C}_\ell^{-1}}\frac{\partial {\bf C_\ell}}{\partial
p_\beta} {\bf \tilde{C}_\ell^{-1}} \right) \ , \label{eq:Fisher} \ee
where $p_{\alpha(\beta)}$ is the $\alpha(\beta)$th cosmological
parameter and ${\bf \tilde{C}_\ell}$ is the ``observed'' covariance
matrix with elements $\tilde{C}^{XY}_\ell$ that include
contributions from noise: \be \tilde{C}^{XY}_\ell=
C^{XY}_\ell+N^{XY}_\ell \ . \label{eq:NoiseAdd} \ee The
expression~(\ref{eq:Fisher}) assumes that all fields $X(\hat{\bf
n})$ are measured over contiguous regions covering a fraction
$f_{\rm sky}$ of the sky. The value of the lowest multipole can be
estimated from $\ell_{\rm min} \approx [\pi /(2f_{\rm sky})]$, where
the square brackets denote the rounded integer.

In general, the noise matrix $N^{XY}_\ell$ receives the
contributions from both the statistical and the systematic errors.
For the statistical error, we assume the uncorrelated Poisson noise
on the galaxy overdensity in each galaxy bin ($G_i$) and shear
fields ($\epsilon_i$), the noise is given by~\cite{HuJain}
\begin{eqnarray}
&&N^{\epsilon_i \epsilon_j}_\ell = \delta_{ij} \frac{\gamma^2_{\rm rms}}{n_j}   \nonumber \\
&&N^{G_i G_j}_\ell = \delta_{ij} \frac{1}{n_j}  \nonumber \\
&&N^{G_i \epsilon_j}_\ell = 0, \label{eq:Noise}
\end{eqnarray}
where $\gamma_{\rm rms}$ is the expected root mean square shear of
the galaxies, and $n_j$ is the number of galaxies per steradian in
the $j$th redshift bin.

Besides the shot noise, we follow \cite{Huterer:2005ez} and consider
three classes of the systematics: redshift errors, additive errors
and multiplicative errors~\footnote{Systematics are notoriously
difficult to model and predict for the forecast, and it can also
stem from some physical processes such as the baryonic uncertainty
studied in~\cite{Zentner:2007bn}. But we didn't include this effect
for simplity.}. The redshift errors may stem from three sources: the
distortion of the total galaxy distribution, $z$-bias and
$z$-scatter. In our calculation, we marginalized over 30 Chebyshev
expansion coefficients describing the distortion to the overall
shape of the galaxy distribution; one redshift bias parameter and
one redshift scatter parameter for each redshift bin to account for
the uncertainty of the redshift measurement of the bins. The
additive error can be generated by the anisotropy of the \emph{Point
Spread Function} (PSF) and they generally present for both galaxy
counts and lensing shear bins. Following \cite{Huterer:2005ez}, we
parametrize the additive error as \be\label{add} (C_\ell^{\rm
XY})_{ij}=\delta_{\rm XY}{\rho}A_i^{\rm X}A_j^{\rm
Y}\Big(\frac{\ell}{\ell_{\ast}^{\rm X}}\Big)^\eta \ee and choose
$\rho=1, \eta=0$. The fiducial values of the $A$'s are chosen to be
conservative, $(A^{\rm g}_i)^2=10^{-8}, (A^{\gamma}_i)^2=10^{-9}$.
The multiplicative errors in measuring the shear can be
parameterized as, \be\label{mul} (\widetilde{C}_{\ell}^\gamma)_{ij}=
(C_{\ell}^\gamma)_{ij}[1+f_i+f_j].\ee

\begin{table*}[thb]
\caption{The $\chi^2$, $\chi_{\rm red}^2$ (reduced $\chi^2$) and
AIC, BIC values for various models. For each quantity, the
difference with respect to the base model $\Lambda$CDM is also
listed.}
 \vspace{0.5cm}
\begin{tabular*}{0.9\textwidth}{@{\extracolsep{\fill}} lcccc cccc}
\hline \hline
Model                &$\chi^2$ &$\Delta\chi^2$ &$\chi_{\rm red}^2(\times10^3)$  &$\Delta\chi_{\rm red}^2(\times10^3)$& BIC    &$\Delta$BIC  & AIC  &$\Delta$AIC \\
\hline
 $\Lambda$CDM        &3149.1   &0              &1057.0                          &0                                   &3197.1  &0            &3161.1      &0            \\
 $N=1$               &3147.8   &$-1.3$         &1056.9                          &$-0.1$                              &3203.9  &6.8          &3161.9      &0.8            \\
 $N=2$               &3146.8   &$-2.3$         &1056.9                          &$-0.1$                              &3210.8  &13.7         &3162.8      &1.7            \\
 $N=3$               &3146.2   &$-2.9$         &1056.9                          &$-0.1$                              &3218.2  &21.1         &3164.2      &3.1            \\
 $N=4$               &3143.3   &$-5.8$         &1056.5                          &$-0.5$                              &3223.3  &26.2         &3163.3      &2.2            \\
 $N=5$               &3139.7   &$-9.4$         &1055.6                          &$-1.4$                              &3227.7  &30.6         &3161.7      &0.6            \\
 $N=6$               &3139.5   &$-9.6$         &1056.0                          &$-1.0$                              &3235.6  &38.4         &3163.5      &2.4            \\
 $N=7$               &3139.4   &$-9.7$         &1056.3                          &$-0.7$                              &3243.4  &46.3         &3165.4      &4.3            \\
 $N=8$               &3139.3   &$-9.8$         &1056.6                          &$-0.4$                              &3251.3  &54.2         &3167.3      &6.2            \\
 $N=9$               &3139.2   &$-9.9$         &1056.9                          &$-0.1$                              &3259.3  &62.2         &3169.2      &8.1            \\
 $N=10$              &3139.2   &$-9.9$         &1057.6                          &$0.6$                               &3267.2  &70.1         &3171.2      &10.1            \\
 $\{w_0,w_a\}$       &3146.7   &$-2.4$         &1056.9                          &$-0.1$                              &3210.7  &13.6         &3162.7      &1.6            \\
  \hline
  \hline
\end{tabular*}
\label{tab:models}
\end{table*}

\subsubsection{Experiments and Cosmological Parameters} \label{sec:experiments}

As mentioned before, the data considered in our forecast include the
SNe observations, CMB power spectra of temperature and polarization
(T and E), WL, GC, and their cross-correlations. We assume CMB T and
E data from the Planck satellite~\cite{planck}, the galaxy
catalogues and WL data by the LSST~\cite{LSST}, complemented by a
futuristic SNe data set provided by a future JDEM space mission. We
strictly follow \cite{Zhao:2008bn} to set up the the survey
parameters. In our forecasts, we use the best fit values of the
cosmological parameters obtained from current data as a fiducial
model, and impose a Gaussian prior on the value of $h$ from the
Hubble Space Telescope (HST)~\cite{HST}. We assign a constant bias
parameter for each galaxy bin, and then marginalize over.

\section{Results} \label{result}

Before viewing the results for parametrization $\mathcal{X_{\rm
I}}$, one might be able to make an `intuitive guess' based on the
following reasoning. As elaborated in Sec. \ref{method}, we attempt
to `see' the possible dark energy dynamics by placing numerous $w$
bins around $-1$ to fit data. If the real dark energy EoS were $-1$,
then the number of bins (and even the method of binning) shouldn't
affect the fit much, i.e., the $\Delta\chi^2$ would have little
dependence on the number of $w$ bins\footnote{In the limit of
infinitely many SNe on the hubble diagram, the $\Delta\chi^2$ is
independent of the binning if the underlying physical model of dark
energy is cosmological constant.}. However, if dark energy were
dynamical, e.g., there exist some local features in $w(z)$ on some
scale, the $w$ bins wander around $-1$ trying to find a better fit
to data than $w=-1$, and the $\Delta\chi^2$ behavior would be
strongly depend upon the number of $w$ bins. Consider, if the
binning is coarse, then it might not be able to resolve the features
in $w(z)$ properly, and this might result in the marginal
improvement on $\chi^2$. However, if the binning resolution is fine
enough allowing $w(z)$ to vary on the scale of main feature of
$w(z)$, the bins may gain sufficient freedom to capture the
principal features in the data set, giving rise to sharp drops on
$\chi^2$. It is true that using finer binning can in principle
improve $\chi^2$ further, but it is not effective to improve the
reduced $\chi^2$, which is an indicator of the goodness of fit, and
is defined as, \be\label{eq:chi2red} \chi^2_{\rm
red}\equiv\chi^2/\nu\ee where $\nu$ denotes the number of degrees of
freedom in the fit, i.e., the number of data points subtracted by
the number of fit parameters defined in Eq.~(\ref{eq:paratriz}). For
a reliable fit, the $\chi^2_{\rm red}$ should be close to unity,
which is the case for all our $10$ fits shown in Fig~\ref{fig:wz}.
For example, for our base model ($\Lambda$CDM), the $\chi^2$ is
$3149.06$ for $2978$ degrees of freedom, giving $\chi^2_{\rm
red}=1.057$.

Using a unnecessary large number of bins can hardly improve the fit.
On the one hand, the super fine bins can do nothing more but resolve
the unimportant feature in the data, or overfit the data by treating
the noise as features, so that they cannot improve the $\chi^2$
drastically. On the other hand, the redundant bins will be highly
correlated, making it difficult to extract useful information from
data by performing the global fit even if the MCMC algorithm is
used. Therefore, there is a trade-off between the number of $w$ bins
$N$ and the GoF, and it is a necessity to seek for the optimal $N$
maximizing the GoF.

Now let's look at the reconstructed $w(z)$ from the current data
shown in Fig~\ref{fig:wz}. With respect to the $\Lambda$CDM model as
a base model, we plot the improvement of the $\chi^2$ in panel (A)
of Fig~\ref{fig:chi2} and in Table \ref{tab:models}. We find that as
$N$ increases, the $\chi^2$ generally decreases, showing an
improvement of the fit with more $w$ bins which is expected.
Furthermore, the slope of $\Delta\chi^2$ varies with $N$, namely,
when $N$ goes from $3$ to $5$, the $\chi^2$ drops quickly whereas
$N<3$ or $N>5$, the $\chi^2$ decreases slightly. As $N$ increases
from $1$ to $3$, $w(z)$ starts to deviate from $-1$. But this makes
the $\chi^2$ get only slightly improved, meaning that these binnings
are too coarse to capture the features properly. When $N$ grows to
$4$, a new pattern appears, say, all the $w$ bins show deviations
from $-1$ at $68\%$ confidence level, and interestingly, the odd bin
values become greater than $-1$ whereas the even bins drop below the
$-1$ boundary. Compared to the case of $N=3$, the resolution of
$N=4$ allows us to see a new feature, say, $w(z)<-1
~(z\in[0.33,0.66])$. This new feature might explain the `drop' on
the $\Delta\chi^2$ plot. As $N$ reaches $5$, the pattern of $w(z)$
resembles that of $N=4$, namely, \ba\label{eq:ineq} && w(z)>-1,
~~z{\in}\mathcal{S}=[0,0.2]\cup[0.4,1.0];\nonumber  \\
 && w(z)<-1,~~z{\in}\mathcal{\bar{S}}. \ea
but the dynamics is seemingly more pronounced -- the deviations from
$-1$ of the first three bins are enhanced to about $95\%$ confidence
level. This makes the $\Delta\chi^2$ take a nosedive again.  When
$N$ exceeds 5, the $w(z)$ pattern remains, but the $\chi^2$ gets
improved only marginally since all the main features have already
been identified by $5$ bins and there is little work left for the
extra bins to do. This is in consistency with our foregoing
`intuitive guess' for the case of dark energy with dynamics.

To quantify the GoF for different $N$, we need to view the reduced
$\chi^2$ versus the number of $w$ bins $N$ shown in panel (B) of
Fig~\ref{fig:chi2} and in Table \ref{tab:models}. As we see, the
improved $\chi^2_{\rm red}$ reaches its extremum at $N=5$. This
means that using five bins is sufficient to find all the important
features on $w(z)$, and any extra bins are redundant. They dilute
the constraints by introducing degeneracies, but can do little to
improve the fit. Therefore $N=5$ is the optimal number of bins we
need.

\begin{table}
\caption{The mean values of the dark energy parameters with 68\% and
95\% C.L. error bars for $N=5$ model and for the CPL model. For the
$N=5$ model, the constraints on the rotated parameters $q$'s are
also listed.} \vspace{0.5cm}

\begin{tabular}{cc|cc}
\hline \hline
$w_1$               &$-0.79^{+0.12+0.23}_{-0.12-0.24}$      &$q_1$      &$-0.93^{+0.07+0.13}_{-0.07-0.15}$   \\
$w_2$               &$-2.1^{+0.49+0.93}_{-0.51-0.79}$       &$q_2$      &$-1.2^{+0.14+0.26}_{-0.14-0.24}$   \\
$w_3$               &$0.89^{+1.1+1.8}_{-1.0-2.1}$           &$q_3$      &$-0.46^{+0.28+0.50}_{-0.28-0.57}$   \\
$w_4$               &$0.41^{+1.9+2.6}_{-1.8-3.0}$           &$q_4$      &$-0.38^{+0.55+0.83}_{-0.58-1.0}$   \\
$w_5$               &$-1.5^{+0.88+1.3}_{-0.91-1.4}$         &$q_5$      &$-1.1^{+0.37+0.62}_{-0.34-0.46}$   \\
  \hline
$w_0$               &$-0.90^{+0.11+0.23}_{-0.11-0.22}$      &$w_a$      &$-0.24^{+0.56+0.98}_{-0.55-1.2}$   \\
  \hline  \hline
\end{tabular}
\label{tab:5bin}
\end{table}

\begin{figure}[tbp]
\includegraphics[scale=1.9]{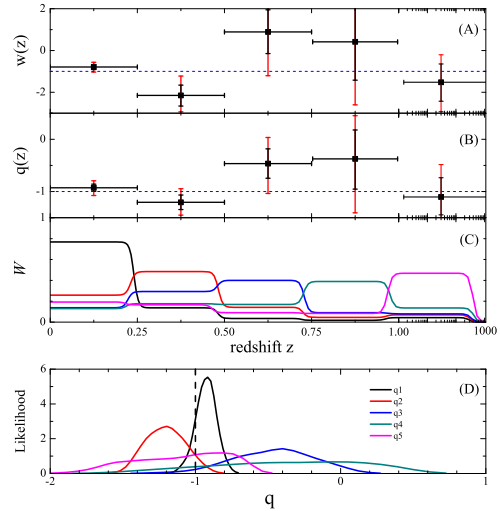}
\caption{ Panel (A,B): The reconstructed $w(z)$ and $q(z)$ using 5
bins from current data. The vertical black and red error bars show
$1\sigma$ and $2\sigma$ errors respectively. The blue dashed line
shows $\Lambda$CDM model; Panel (C): The re-scaled window function
$\mathcal{W}$; Panel (D): 1-D probability distribution of the $q$'s.
}\label{fig:merge_4b}
\end{figure}

Note, however, different model selection criteria may prefer
different models. Here we consider two other widely used
alternatives, the \emph{Akaike information criterion}
(AIC)~\cite{AIC} and the \emph{Bayesian Information Criterion}
(BIC)~\cite{BIC}. They both serve as tools to compare different
models using a likelihood method, yet they base on different
statistical arguments. For example, AIC stems from the minimization
prescription of the Kullback-Leibler information entropy, yet BIC
roots in the approximation of the Bayes factor. The AIC and BIC are
defined as,

\ba\label{ic} &&{\rm AIC}=-2~{\rm ln}~\mathcal{L}+2N_{\rm P} \\
&&{\rm BIC}=-2~{\rm ln}~\mathcal{L}+N_{\rm P}~{\rm ln}N_{\rm D} \ea
where $\mathcal{L}$ is the maximum likelihood, and $N_{\rm P}$ and
$N_{\rm D}$ are the numbers of free parameters and that of the data
points used in the fit, respectively. Viable models are supposed to
minimize the quantities of AIC or BIC. The AIC and BIC for different
models are listed in Table \ref{tab:models}, which reads,

\begin{enumerate}
  \item Both AIC and BIC favor the $\Lambda$CDM model;
  \item Model of $N=5$ is favored by the reduced $\chi^2$ criterion,
  and is mildly disfavored by AIC;
  \item All models except $\Lambda$CDM are strongly disfavored by
  BIC.
\end{enumerate}

It is not surprising that AIC and BIC resist dynamical dark energy
models because both AIC and BIC include strong penalty terms
inhabiting the overfit -- reducing $\chi^2$ by introducing redundant
free parameters. And if the number of data points exceeds
$e^2\sim7$, which is often the case in cosmology, the penalty of BIC
for additional parameters is stronger than that of the AIC. To be
explicit, let's suppose one model has $\Delta{N_{\rm P}}$ additional
free parameters compared to $\Lambda$CDM. If these $\Delta{N_{\rm
P}}$ extra parameters can help reduce $\chi^2$ by $2\Delta{N_{\rm
P}}$, then it is preferred to $\Lambda$CDM by AIC. However, to
satisfy BIC, the $\chi^2$ must be reduced by three times more, i.e.
$\Delta\chi^2\sim8\Delta{N_{\rm P}}$.

However, no matter favored or not by AIC or BIC, the dynamical dark
energy model for the $N=5$ case is worth investigating in depth
since it reduces the $\chi^2$ most for each degree of freedom, and
the reconstructed $w(z)$ shows an excellent convergence.
\begin{figure*}
\includegraphics[scale=2.5]{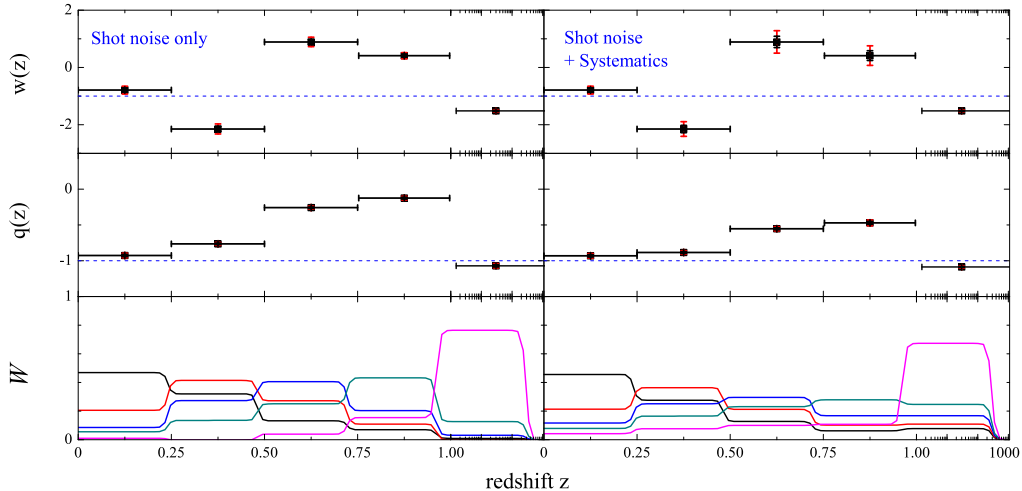}
\caption{Future forecasts from JDEM+Planck+LSST on $w(z),q(z)$ and
the window functions. Left panels: Forecasts including statistical
errors only; Right: Statistical plus systematic errors. The legends
are the same as that for panels (A-C) in Fig~\ref{fig:merge_4b}.
}\label{fig:LSST}
\end{figure*}
But unfortunately, this reconstruction is blurred by the
correlations among all the $w$ bins, making Fig~\ref{fig:wz} and
Eq.~(\ref{eq:ineq}) hard to interpret. The correlation between bins
$w_i$ and $w_j$ is defined as,

\be\label{eq:corr} {\rm Corr}(w_i,w_j)=\frac{{\rm
Cov}(w_i,w_j)}{\sigma(w_i)\sigma(w_j)}. \ee where ${\rm
Cov}(w_i,w_j)$ is the covariance between bins $w_i$ and $w_j$, and
$\sigma(w_i)$ and $\sigma(w_j)$ are standard deviations of $w_i$ and
$w_j$ respectively. In Fig~\ref{fig:corr}, we plot the correlation
matrices for the cases of $N=2$ to $N=10$. As we can see, for the
$w$ bins centering at $z\lesssim0.5$, there exists strong
correlation between the neighboring bins. The last bin
($z\in[1,1000]$) has little correlation with other bins, which
explains why the constraint on this bin doesn't get diluted even if
we have large number of redundant bins, as shown in
Fig~\ref{fig:wz}.

To de-correlate the bins, we apply the LPCA procedure explained in
Sec.~\ref{method} to rotate the $w$'s into $q$'s, and we find, \be
\label{eq:qz} q(z)\left\{
    \begin{array}{ll}
    >-1~(68\%~C.L.),~z\in[0,0.25); \\
    <-1~(\sim95\%~C.L.),~z\in[0.25,0.5); \\
    >-1~(\sim95\%~C.L.),~z\in[0.5,0.75); \\
    >-1~(68\%~C.L.),~z\in[0.75,1); \\
    \in[-1.44,-0.73]~(68\%~C.L.),~z\geq1. \\
    \end{array}
\right. \ee

So at low redshift $z<0.25$ or at high redshift $z\geq0.75$, $q(z)$
is consistent with the $\Lambda$CDM prediction at the 95\%
confidence level. However, in the redshift range [0.25, 0.75),
$q(z)$ crosses the cosmological constant boundary, and the error
bars on the $q$'s are uncorrelated thus free of degeneracy by
design.

The detailed result of the LPCA is summarized in Table
\ref{tab:5bin} and in Fig~\ref{fig:merge_4b}. From Panel (C) in
Fig~\ref{fig:merge_4b}, we see that the window functions are almost
positive and fairly localized, making the $q$'s directly relate to
the $w$'s. This means that the $q$'s have almost one-to-one
correspondence with the original $w$ bins. Therefore we can come to
conclusion that the $\Lambda$CDM model is generally consistent with
current data, yet Eq.~(\ref{eq:qz}) implies some hint, very weak
though, on the possible dynamics of dark energy. In particular, a
model predicting $w(z)<-1$ if $z\in[0.25,0.5)$ and $w(z)>-1$ if
$z\in[0.5,0.75)$, which means that $w(z)$ crosses $-1$ in the range
of $z\in[0.25,0.75)$, is mildly favored.

\begin{figure*}
\includegraphics[scale=1.6]{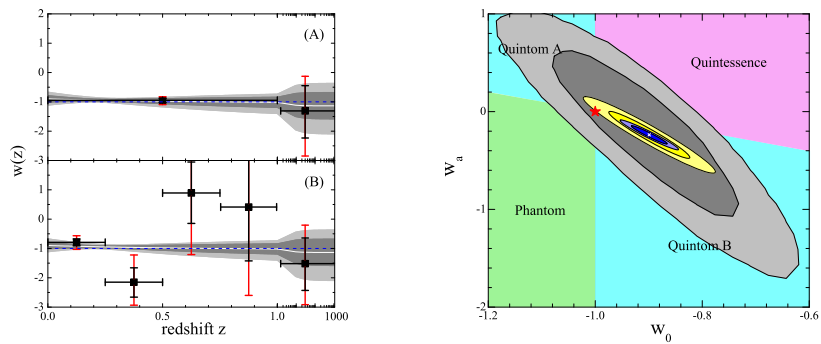}
\caption{Left: Comparison of the reconstructed $w(z)$ using two
different parametrizations. The 68\%(black) and 95\%(red) C.L. error
bars show the reconstructed $w(z)$ using parametrization
$\mathcal{X_{\rm I}}$; The shaded regions show the $w(z)$
reconstruction using parametrization $\mathcal{X_{\rm II}}$. The
inner and outer shades illustrate the $68\%$ and $95\%$ C.L. errors
respectively, and central white line show the best fit model; Right:
Contour plots for $w_0,~w_a$. Grey: current data; Yellow: forecast
from JDEM+Planck+LSST with systematics; Blue: forecast without
systematics. For the same color, the dark and light shaded regions
denote the $68\%$ and $95\%$ C.L. contours respectively. The red
star stands for the $\Lambda$CDM model, and the white cross
illustrates the current best fit model. }\label{fig:compare}
\end{figure*}

Based on the best fit $w(z)$ from current data as the fiducial
model, we can make the future forecast from JDEM, in combination of
the Planck and LSST surveys. The results are shown in
Fig~\ref{fig:LSST}. As we can see, the window functions are
localized as that for the current data, and the error bars on $w(z)$
get shrink by roughly a factor of $8$ and $13$, for the cases with
systematics and without systematics, respectively. Note that
including systematics changes the correlations among the dark energy
bins, which gives rise to the corresponding changes in $\mathcal{W}$
and the $q$'s. Given the constraints on the $q$'s, we can conclude
that, if the fiducial model we derived from current data were true,
then even for the most conservative case where the full systematics
are included, JDEM combined with Planck and LSST would be able to
detect a roughly $10\sigma$ deviation of $w(z)$ from $-1$.

It is useful to study the dependence of the results on the form of
dark energy parametrizations. For an illustration, we compare the
result we obtained so far to that from parametrization
$\mathcal{X_{\rm II}}$, i.e., the CPL parametrization, and the
result is summarized in Table \ref{tab:models} and in
Fig~\ref{fig:compare}. As shown, we find that the constraints on the
CPL parameters from current data are \be\label{eq:w0wa_constraint}
w_0=-0.90^{+0.11+0.23}_{-0.11-0.22},~w_a=-0.24^{+0.56+0.98}_{-0.55-1.2}.
\ee The central values indicate that the `quintom B' scenario is
mildly favored, namely, the EoS today $w(z)|_{z=0}=w_0>-1$, while
EoS in the far past $w(z)|_{z=\infty}=w_0+w_a<-1$. This is
consistent with the recent published result using the `Constitution'
SNe sample~\cite{CfA_fit}. The contour plots of $w_0$ and $w_a$ are
shown as shaded grey regions in the right panel of
Fig~\ref{fig:compare}. We can see the best fit model from current
data lies within the `quintom B' region, and the $\Lambda$CDM model
is consistent with current data at $1\sigma$. Based on the current
best fit model, future surveys has the ability to exclude
$\Lambda$CDM model at $2\sigma$ and $5\sigma$ for the cases with and
without the systematics respectively. In the left panels, we make a
direct comparison of the reconstructed $w(z)$ from two different
parametrizations -- binned $w$ and the CPL, shown in error bars and
shaded regions respectively. Interestingly, we find that the $N=2$
result agrees well with CPL (and the $\chi^2$ for these two fits are
very similar, see Table \ref{tab:models}), whereas for the case of
$N=5$, there is an apparent discrepancy. This is as expected -- the
results converge if the numbers of free parameters in the fit are
the same, however, the CPL parametrization does not have enough
freedom to resolve the local details of $w(z)$, resulting in the
failure to capture the `dip' and `bump' happening at
$0.25\lesssim{z}\lesssim0.75$.

\section{Discussion and Summary}\label{conclusion}

In this work, we have investigated the constraints on the general
form of the equation-of-state of dark energy from the latest SNe,
CMB and LSS data. We utilize a model-independent strategy --
redshift binning plus PCA -- to extract information from data as
much as possible. Starting from the most general parametrization --
the $w$ binning using the smooth tanh bins, we constrain the $w$
bins using a MCMC algorithm while paying particular attention to the
consistent implementation of the dark energy perturbations. We
repeat this procedure for different binning scheme and investigate
the goodness of fits using three different model selecting criteria
-- the reduced $\chi^2$, AIC and BIC. While AIC and BIC strongly
favor the $\Lambda$CDM model, the 5-$w$-bin model gives the maximum
reduced $\chi^2$, and we find a convergent evolution trend of $w(z)$
when $N\geq4$.

We choose the model of $N=5$ and rotate the $w$ bins into the $q$
bins to eradicate the correlations using a LPCA method, and we found
that at low redshift $z<0.25$ or at high redshift $z\geq0.75$,
$q(z)$ is consistent with $-1$ at the 95\% confidence level.
However, in the intermediate redshift range [0.25, 0.75), $q(z)$
crosses the cosmological constant boundary. Since the $q$'s have
almost one-to-one correspondence with the original $w$ bins by
design, we can draw the conclusion that the $\Lambda$CDM model is
generally consistent with the current data, yet there exists some
weak hint of the possible dynamics of dark energy. In particular, a
quintom model predicting $w(z)<-1$ if $z\in[0.25,0.5)$ and $w(z)>-1$
if $z\in[0.5,0.75)$, which means that $w(z)$ crosses $-1$ in the
range of $z\in[0.25,0.75)$, is mildly favored.

Note that the measurement of the luminosity distance is crucial to
study dark energy dynamics, thus our results are sensitive to the
choice of the SN data set and analysis. The apparent dynamics of
dark energy we found using the published `Constitution' SNe sample
might be physical, but there is another possibility of the artifacts
in the SNe data analysis. Therefore we have planned a careful study
of the effect of SN sample on dark energy constraints.

Given the best fit model from current data, we make a forecast from
the upcoming/future surveys of JDEM, Planck and LSST, and we find
that the future data are able to shrink the error bars on the dark
energy bins by roughly a factor of $10$, which is promising to find
the smoking gun of the dark energy evolution.

Note added: After completion of this work, we noticed another
analysis on $w(z)$ using different dataset and
method~\cite{Serra:2009yp}. Although both our work and
Ref.~\cite{Serra:2009yp} claim that the cosmological constant is
favored at 95\% CL., Ref.~\cite{Serra:2009yp} found less preference
of dark energy dynamics. This difference may stem from the
difference in data selection and analyzing prescription, and it is
worth further investigating.

\acknowledgments

All of our numerical calculations were preformed on WestGrid in
Canada. It's a pleasure to thank Zuhui Fan, Dragan Huterer, Eiichiro
Komatsu, Hong Li, Eric Linder, Jie Liu, Levon Pogosian, Jun-Qing
Xia, Hu Zhan and Joel Zylberberg for helpful discussions. We also
would like to thank the anonymous referees for insightful comments
and suggestions. GZ is supported by NSERC, funds from Simon Fraser
University and from the European Research Council, and XZ by
National Science Foundation of China under Grant Nos. 10533010 and
10675136, by the 973 program No. 2007CB815401, and by the Chinese
Academy of Science under Grant No. KJCX3-SYW-N2.


\end{document}